\documentclass[prd,twocolumn,amsmath,amssymb,floatfix,superscriptaddress]{revtex4}

\usepackage{bm}
\usepackage{amsmath}
\usepackage{epsf}
\usepackage{color}
\usepackage{natbib}
\usepackage{graphicx}
\usepackage{hyperref}
\usepackage{ifthen}

\newcommand{\lsc}{\mathcal{L}}
\newcommand{\dlnl}{-2\Delta\ln\lsc}

\newcommand{\zmax}{z_{\rm max}}
\newcommand{\ok}{\Omega_{\rm K}}
\newcommand{\simlt}{\lesssim}

\begin{document}
\title{Evidence for horizon-scale power from CMB polarization}

\author{Michael J. Mortonson}\email{mjmort@uchicago.edu}
\affiliation{Department of Physics, University of Chicago, Chicago IL 60637}
\affiliation{Kavli Institute for Cosmological Physics and Enrico Fermi Institute, University of Chicago, Chicago IL 60637, U.S.A.}

\author{Wayne Hu}
\affiliation{Kavli Institute for Cosmological Physics and Enrico Fermi Institute, University of Chicago, Chicago IL 60637, U.S.A.}
\affiliation{Department of Astronomy \& Astrophysics, University of Chicago, Chicago IL 60637}

\date{\today}

\begin{abstract}
The CMB temperature power spectrum offers
ambiguous evidence for the existence of horizon-scale power in the primordial power 
spectrum
due to  uncertainties in spatial curvature and the physics of cosmic acceleration as well as the observed low quadrupole.
Current polarization data from WMAP provide evidence
for horizon-scale power that is robust to these uncertainties. 
Polarization on the largest scales arises mainly from scattering at $z\simlt 6$ when
the universe is fully ionized, making the evidence robust to ionization history variations at higher redshifts as well.
A cutoff in the power spectrum 
is limited to $C= k_{C}/10^{-4}$ Mpc$^{-1}<5.2$ (95\% CL) by polarization, only 
slightly
weaker than joint temperature and polarization constraints in flat $\Lambda$CDM
($C<4.2$).
Planck should improve the polarization limit to $C<3.6$ for any model of the acceleration epoch and ionization history as well
as provide tests for foreground and systematic contamination.
\end{abstract}

\maketitle

\section{Introduction} \label{sec:intro}

Whether or not  the CMB temperature power spectrum provides
evidence for horizon-scale power in the primordial power spectrum
 depends on assumptions about spatial curvature and the
physics of 
late-time cosmic acceleration.  In the flat $\Lambda$CDM cosmology, roughly
half of the power at the largest angular scales is contributed by the integrated
Sachs-Wolfe (ISW) effect from the decay of the potential during the acceleration
epoch.  Due to projection effects, these contributions primarily come
from fluctuations on scales smaller than a tenth of the current horizon.

Indeed, the large ISW effect presents a challenge for explanations of the observed low CMB temperature quadrupole
in terms of the primordial power spectrum  \cite{Bennett:1996ce,Bennett:2003bz,Spergel:2003cb}.  Nevertheless, this feature
has motivated many studies 
of models in which the usual, nearly scale-invariant, inflationary 
power spectrum is modified by a cutoff that suppresses large-scale 
power \cite{Jing:1994jw,Yokoyama:1998rw,Linde:2003hc,Contaldi:2003zv,Bridle:2003sa,Efstathiou:2003hk,Cline:2003ve,Feng:2003zua,Piao:2003zm,Niarchou:2003hz,Lasenby:2003ur,Kesden:2003zm,Sinha:2005mn,Bridges:2005br,Bridges:2006zm,Boyanovsky:2006pm,Spergel:2006hy,Jain:2008dw}.  In a flat $\Lambda$CDM context,
such models can marginally improve the fit to temperature data 
over that of a power-law spectrum by removing power on 
scales $k\lesssim 3\times 10^{-4}~{\rm Mpc}^{-1}$.  The ISW effect prevents
a more substantial improvement from a horizon-scale cutoff.
Models that suppress power on smaller scales are disfavored by 
the data since the observed power in CMB temperature at $\ell\gtrsim 4$ is 
consistent with the predictions of the power-law spectrum.

These CMB temperature-based conclusions depend strongly on the 
model for late time acceleration.  For example, variations in the equation of
state of the dark energy, spatial curvature, and dark energy clustering
can change the ISW contributions
at low multipoles \cite{Hu:1998kj,Bean:2003fb}.
In more exotic modified gravity models, a horizon-scale cutoff can actually
be strongly favored.  In the self-accelerating braneworld 
model \cite{dvali:00}, a cutoff at $k \sim 8 \times 10^{-4}~{\rm Mpc}^{-1}$ is 
 preferred by the temperature data \cite{Fanetal08}. 

CMB polarization, on the other hand, is free of the ISW effect but 
retains the sensitivity to a large-scale cutoff \cite{Dore:2003wp,Skordis:2004xr,GorHu04,Cline:2003ve}. 
In this paper, we demonstrate the robustness of polarization inferences
about horizon-scale cutoffs in power by comparing the constraints from temperature 
and polarization using both current (5-year) CMB data from WMAP
\cite{Nolta:2008ih,Dunkley:2008ie,Komatsu:2008hk}
and forecasts for data from Planck \cite{Planck:2006uk}.
We describe the fiducial cutoff model and 
 illustrate the degeneracies with the
ISW effect in Sec.~\ref{sec:horizon}. 
We show how CMB polarization constraints on the cutoff model 
are free from these degeneracies as well as  
ionization history uncertainties 
in Sec.~\ref{sec:pol}, and we discuss our conclusions
in Sec.~\ref{sec:discuss}.

\section{Horizon-Scale Power}
\label{sec:horizon}

For definiteness, 
we model the primordial curvature power spectrum with 
an exponential cutoff at $k<k_C$,
\begin{equation}
\Delta_\zeta^2 = A_s \left[1-e^{-(k/k_C)^\alpha}\right] \left( \frac{k}{k_0} \right)^{n_s-1},
\label{eq:cutoff}
\end{equation}
motivated by a transition between stages of inflation \cite{Contaldi:2003zv}.
Here $A_s$ is the normalization of the power spectrum at a pivot scale 
of $k_0 =0.05$ Mpc$^{-1}$ and $n_s$ is the spectral tilt.
We take $\alpha=3.35$ as in Ref.~\cite{Contaldi:2003zv}, but the 
CMB power spectra are insensitive to the exact value of this parameter 
\cite{Sinha:2005mn}.
For notational convenience, we define
\begin{equation}
k_C = C \times 10^{-4}~{\rm Mpc}^{-1} \approx 0.3C H_0/h \,,
\end{equation} 
such that $C$ is of order unity for cutoff scales on the horizon.

\begin{figure}[tb]
  \resizebox{3.5in}{!}{\includegraphics[angle=0]{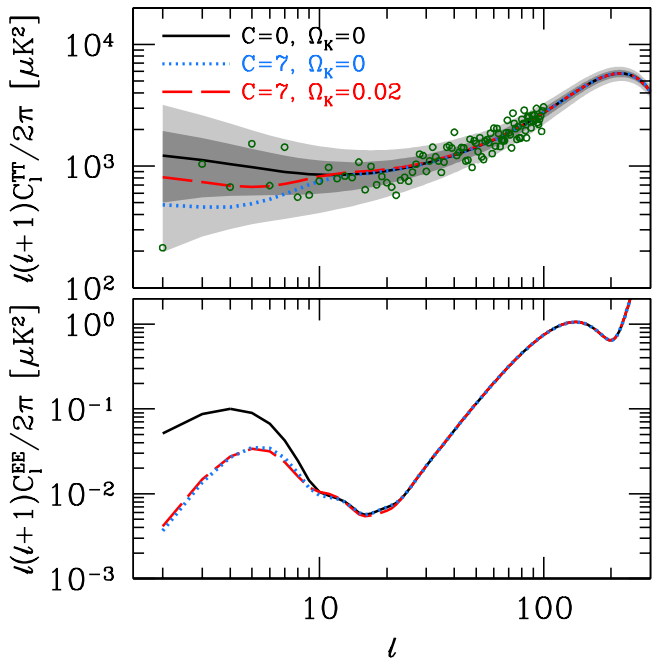}}
  \caption{
The cutoff-curvature degeneracy in the CMB $TT$ 
spectrum (\emph{top}) is broken in the $EE$ spectrum 
(\emph{bottom}). Solid curves show a flat model with 
no large-scale cutoff ($C=0$, with 68\% and 95\% CL 
cosmic variance bands per $\ell$ for $TT$); 
other curves have a cutoff at 
$C=7$.  Open circles show WMAP temperature data at $\ell\leq 100$. 
The WMAP measurement errors are not plotted since they 
are smaller than the point symbols.
For the flat models, $h=0.724$, while the 
open model with $\ok=0.02$ has $h=0.90$ to preserve the CMB acoustic 
scale. All other parameters are the same for the 3 models.
  }
  \label{fig:okkc}
\end{figure}

A reduction of large scale power from a finite cutoff $C>0$ can be partially compensated
by the ISW effect from the late-time decay of gravitational potentials.
For example, negative spatial curvature can substantially boost power at large
angular scales. 
Figure~\ref{fig:okkc} illustrates the degeneracy between 
the scale of the exponential cutoff and 
spatial curvature through the ISW effect.   The polarization spectrum, on the
other hand, does not receive contributions from the ISW effect and better
reveals the presence of a cutoff.   

With WMAP data, the flat model with a cutoff at $C=7$ 
in Fig.~\ref{fig:okkc} can be distinguished from the flat $C=0$ model 
at high significance using temperature data alone.   The likelihood
ratio statistic gives $\dlnl_{TT}=11.7$ relative to the model 
with no cutoff.  However, the open model with a cutoff 
is more difficult to distinguish from the flat $C=0$ model 
using only temperature ($\dlnl_{TT}=3.1$).   With the addition of polarization
data, the open model becomes distinguishable with $\dlnl_{\rm tot}=12.6$.
This example suggests that even with current data, polarization can provide
comparable constraints to temperature in a manner that is robust to 
curvature and the 
physics of the acceleration epoch.

\section{Robust  Polarization Constraints} \label{sec:pol}

\begin{figure}[tb]
  \resizebox{3.4in}{!}{\includegraphics[angle=0]{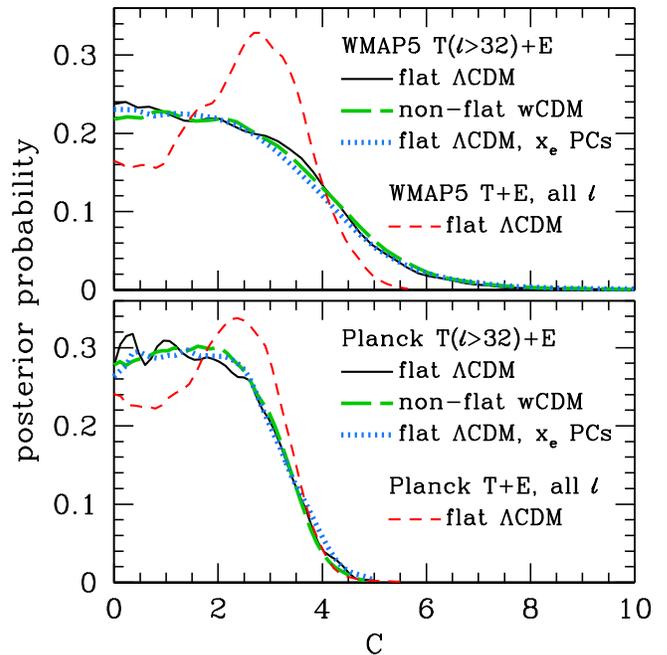}}
  \caption{
Marginalized posterior probability for the 
cutoff scale $C$ using WMAP (\emph{top panel}) and 
simulated Planck data (\emph{bottom panel}), 
showing robustness of
the polarization constraint to curvature, dark energy, and 
the ionization history. 
For Planck, the simulated  
spectra are constrained to the WMAP temperature measurements at 
$\ell\leq 100$. (See text for details.)
}
  \label{fig:kcpostprob}
\end{figure}

To quantify the constraints on horizon-scale power from polarization, test
their robustness, and examine their potential for future measurements, 
 we adopt a Markov Chain Monte Carlo (MCMC) approach. 
We perform this analysis using modified versions of CAMB \cite{Lewis:1999bs}, 
CosmoMC \cite{Lewis:2002ah,cosmomc_url}, 
and the 5-year WMAP likelihood code \cite{Dunkley:2008ie,wmaplike_url}.
The resulting constraints on the cutoff scale $C$ are summarized 
in Fig.~\ref{fig:kcpostprob}.

We begin with an analysis in the flat $\Lambda$CDM cosmology where ISW
contributions are nearly fixed by parameters that are well-determined by
the acoustic peaks.   We add $C$ to the standard set of parameters for 
MCMC analyses of CMB data, $\{\Omega_bh^2,\Omega_ch^2,h,\tau,A_s,n_s\}$.
As expected, with WMAP data 
there is a marginal preference for $C \approx 3$ due to the low observed power in the temperature spectrum at low multipoles
 (see Fig.~\ref{fig:kcpostprob}, top panel)
\cite{Jing:1994jw,Bridle:2003sa,Contaldi:2003zv,Cline:2003ve,Piao:2003zm,Niarchou:2003hz,Sinha:2005mn,Bridges:2005br,Bridges:2006zm,Spergel:2006hy}.
With a flat prior on $C$, the 95\% CL upper bound is $C<4.2$.  Without polarization
data, this constraint weakens to $C<5.3$.

In the flat $\Lambda$CDM context, most of the temperature-only constraint 
on $C$ comes from the temperature
spectrum at $\ell \simlt 30$.   Consequently, the bound is sensitive to assumptions about
the curvature and the dark energy.   To test this dependence, we omit the
temperature data at $\ell \le 32$.
(This is a convenient dividing point since the WMAP likelihood code 
uses different methods to compute the likelihood at scales 
above and below $\ell=32$~\cite{Dunkley:2008ie}.)
With only the temperature data at smaller scales, 
the upper limit is $C \lesssim 25$, reflecting a near elimination of the
constraint.  Nonetheless, once polarization is added the bound
is only marginally weaker than the full (all $\ell$) constraint, 
$C \lesssim 5.2$  (see Fig.~\ref{fig:kcpostprob}, top panel).  
Thus the polarization constraint is 
already competitive with the limits from temperature even in the most restrictive
flat $\Lambda$CDM context.
Moreover, the polarization constraint is less model-dependent:
in Fig.~\ref{fig:kcpostprob}, the posterior probability of $C$ is nearly unchanged
even if both  curvature and  a constant dark energy equation of state
$w$ (with prior $-2<w<0$) are included as additional MCMC parameters and 
marginalized.   In fact, the constraint from polarization
is robust to even more extreme changes such as modified gravity explanations for
cosmic acceleration \cite{Fanetal08}.

Polarization constraints are also robust to uncertainties in the ionization
history.   As noted in Ref.~\cite{Mortonson:2008rx}, predictions for the lowest
few multipoles in polarization are robust to reionization variation
since their contributions arise mainly
from $z\simlt 6$ where we know that the universe is fully ionized.  The impact of
variations at higher redshift from projection effects can be effectively controlled
by measurements at higher multipoles where they make most of their contribution.

To quantify this robustness to the ionization history, we adopt the principal components (PCs)
technique~\cite{Hu:2003gh,Mortonson:2007hq}, in which the 
evolution of the ionization fraction is parametrized as
\begin{equation}
x_e(z) = x_e^{\rm fid}(z) + \sum_{\mu=1}^N m_{\mu} S_{\mu}(z),
\end{equation}
where $x_e^{\rm fid}(z)$ is an arbitrary fiducial model (taken here to 
be constant $x_e^{\rm fid}=0.15$), $S_{\mu}(z)$ are the reionization PCs, 
and the PC amplitudes $m_{\mu}$ 
are subject to physicality bounds corresponding to $0\leq x_e\leq 1$ 
as described in Ref.~\cite{Mortonson:2007hq}.  The PC code 
modifications to CAMB and CosmoMC have been made
publicly available \cite{cambrpc_url}.
For a conservative upper limit to the start of reionization of $\zmax=30$, 
$N=5$ PCs are sufficient to completely represent the effects of 
ionization variation on the CMB polarization power spectrum~\cite{Mortonson:2007hq}.
  Hence
marginalizing these parameters makes constraints on $C$ robust to
{\it any} ionization history for $z<30$. In  
Fig.~\ref{fig:kcpostprob} (top panel), we show that WMAP
constraints on $C$ are almost entirely unchanged by the marginalization 
over reionization parameters.

In the near future, constraints on horizon-scale power should be dominated
by polarization information.    We use a simulated
 temperature and polarization data set to make a forecast for the recently launched
 Planck satellite.   It is important here to account for the fact that WMAP has
 already measured the temperature power spectrum to the cosmic variance limit at low
 multipoles (see Fig.~\ref{fig:okkc})
 and that there is a marginal preference for a finite cutoff.
 We therefore define the simulated temperature power spectrum for Planck as
\begin{equation}
\hat C_\ell^{TT} = 
\begin{cases}
\hat C_\ell^{TT({\rm WMAP})}, & \ell \le 100, \\
C_\ell^{TT({\rm fid})}, & \ell > 100, \\
\end{cases}
\end{equation}
where $\hat C_\ell^{TT({\rm WMAP})}$ is taken from the WMAP data and
 $C_\ell^{TT{(\rm fid)}}$ is the temperature power spectrum of the fiducial
flat $\Lambda$CDM model, 
$\{\Omega_bh^2=0.0224,\Omega_ch^2=0.108,h=0.724,\tau=0.089,A_s=2.137\times 10^{-9},n_s=0.96\}$.
For the polarization data, we take 
the ensemble mean of the fiducial model given the
WMAP temperature constraint
\begin{eqnarray}
\hat C_\ell^{EE} &=& C_\ell^{EE({\rm fid})}  \left[1 + R_\ell^2 \left( \frac{\hat C_\ell^{TT}}{C_\ell^{TT({\rm fid})} } -1 \right)\right] \,, \nonumber\\
\hat C_\ell^{TE} &=& C_\ell^{TE({\rm fid})}  \frac{\hat C_\ell^{TT}}{C_\ell^{TT({\rm fid})} }\,,
\end{eqnarray}
where
\begin{equation}
R_\ell =\frac{C_\ell^{TE({\rm fid})}}{ \sqrt{C_\ell^{TT({\rm fid})} C_\ell^{EE({\rm fid})} }}
\end{equation}
is the temperature-polarization correlation coefficient in the fiducial model.

For the Planck satellite noise specifications, we take a combination of the central
70, 100, and 143 GHz channels with the sensitivity and resolution given 
in Ref.~\cite{Planck:2006uk}.   
The other Planck frequency 
channels are effectively used for foreground monitoring and removal as well as
checks for systematic effects.
The results of the MCMC for Planck are shown in Fig.~\ref{fig:kcpostprob} (bottom panel).  Note
that now the constraints with and without the $\ell \le 32$ multipoles in the temperature
spectrum are comparable, aside from the slight preference for $C \sim 3$ in the former case, 
showing that polarization is expected to dominate Planck's constraint on $C$.
Like the WMAP polarization constraint, 
the upper limit $C\lesssim 3.6$ (95\% CL) from Planck polarization data 
remains robust to marginalization of the curvature and
a constant dark energy equation of state.    It is robust to marginalization
over the PC reionization parameters as well.

\section{Discussion} \label{sec:discuss}

We have shown that current polarization data from WMAP provide evidence
for horizon-scale power that, unlike the evidence from temperature data, 
is robust to uncertainties in spatial curvature,
dark energy, and ionization history variations.  
A cutoff in the inflationary power spectrum 
is limited to $C= k_{c}/10^{-4}$ Mpc$^{-1}<5.2$ (95\% CL) by polarization. 
This constraint is only slightly
weaker than joint temperature and polarization limits in flat $\Lambda$CDM
($C<4.2$).

Data from Planck should improve these constraints to the point where polarization dominates
the limit for any model of the acceleration epoch and ionization history.
Statistical errors should improve to $C<3.6$ if
the true model has no cutoff.  Perhaps more importantly, the larger frequency 
coverage of Planck should make the constraint more robust to foregrounds
and other systematic effects.

\vspace{1cm}
{\it Acknowledgments:}  This work was supported by NSF PHY-0114422 and
 NSF PHY-0551142 at the KICP.  MJM was additionally supported by the NSF GRFP. 
WH was additionally supported by DOE contract DE-FG02-90ER-40560 and
 the  Packard Foundation.

\vfill
\bibliographystyle{arxiv_physrev}
\bibliography{pkcutoff}

\def\eprinttmppp@#1arXiv:@{#1}
\providecommand{\arxivlink[1]}{\href{http://arxiv.org/abs/#1}{arXiv:#1}}
\def\eprinttmp@#1arXiv:#2 [#3]#4@{\ifthenelse{\equal{#3}{x}}{\ifthenelse{
\equal{#1}{}}{\arxivlink{\eprinttmppp@#2@}}{\arxivlink{#1}}}{\arxivlink{#2}
  [#3]}}
\providecommand{\eprintlink}[1]{\eprinttmp@#1arXiv: [x]@}
\renewcommand{\eprint}[1]{\eprintlink{#1}}
\providecommand{\eprintmod}[1][XXXX.XXXX]{\eprintlink{#1}}
\providecommand{\adsurl}[1]{\href{#1}{ADS}}
\renewcommand{\bibinfo}[2]{\ifthenelse{\equal{#1}{isbn}}{\href{http://cosmolog%
ist.info/ISBN/#2}{#2}}{#2}}
\begin{thebibliography}{41}
\expandafter\ifx\csname natexlab\endcsname\relax\def\natexlab#1{#1}\fi
\expandafter\ifx\csname bibnamefont\endcsname\relax
  \def\bibnamefont#1{#1}\fi
\expandafter\ifx\csname bibfnamefont\endcsname\relax
  \def\bibfnamefont#1{#1}\fi
\expandafter\ifx\csname citenamefont\endcsname\relax
  \def\citenamefont#1{#1}\fi
\expandafter\ifx\csname url\endcsname\relax
  \def\url#1{\texttt{#1}}\fi
\expandafter\ifx\csname urlprefix\endcsname\relax\def\urlprefix{URL }\fi

\bibitem{Bennett:1996ce}
C.~L. Bennett {\em et~al.},
\newblock Astrophys. J. {\bf 464}, L1 (1996),
  [\eprintmod[arXiv:astro-ph/9601067]].

\bibitem{Bennett:2003bz}
C.~L. Bennett {\em et~al.},
\newblock Astrophys. J. Suppl. {\bf 148}, 1 (2003),
  [\eprintmod[arXiv:astro-ph/0302207]].

\bibitem{Spergel:2003cb}
D.~N. Spergel {\em et~al.},
\newblock Astrophys. J. Suppl. {\bf 148}, 175 (2003),
  [\eprintmod[arXiv:astro-ph/0302209]].

\bibitem{Jing:1994jw}
Y.-P. Jing and L.-Z. Fang,
\newblock Phys. Rev. Lett. {\bf 73}, 1882 (1994),
  [\eprintmod[arXiv:astro-ph/9409072]].

\bibitem{Yokoyama:1998rw}
J.~Yokoyama,
\newblock Phys. Rev. {\bf D59}, 107303 (1999).

\bibitem{Linde:2003hc}
A.~Linde,
\newblock JCAP {\bf 0305}, 002 (2003), [\eprintmod[arXiv:astro-ph/0303245]].

\bibitem{Contaldi:2003zv}
C.~R. Contaldi, M.~Peloso, L.~Kofman and A.~Linde,
\newblock JCAP {\bf 0307}, 002 (2003), [\eprintmod[arXiv:astro-ph/0303636]].

\bibitem{Bridle:2003sa}
S.~L. Bridle, A.~M. Lewis, J.~Weller and G.~Efstathiou,
\newblock Mon. Not. Roy. Astron. Soc. {\bf 342}, L72 (2003),
  [\eprintmod[arXiv:astro-ph/0302306]].

\bibitem{Efstathiou:2003hk}
G.~Efstathiou,
\newblock Mon. Not. Roy. Astron. Soc. {\bf 343}, L95 (2003),
  [\eprintmod[arXiv:astro-ph/0303127]].

\bibitem{Cline:2003ve}
J.~M. Cline, P.~Crotty and J.~Lesgourgues,
\newblock JCAP {\bf 0309}, 010 (2003), [\eprintmod[arXiv:astro-ph/0304558]].

\bibitem{Feng:2003zua}
B.~Feng and X.~Zhang,
\newblock Phys. Lett. {\bf B570}, 145 (2003),
  [\eprintmod[arXiv:astro-ph/0305020]].

\bibitem{Piao:2003zm}
Y.-S. Piao, B.~Feng and X.-m. Zhang,
\newblock Phys. Rev. {\bf D69}, 103520 (2004),
  [\eprintmod[arXiv:hep-th/0310206]].

\bibitem{Niarchou:2003hz}
A.~Niarchou, A.~H. Jaffe and L.~Pogosian,
\newblock Phys. Rev. {\bf D69}, 063515 (2004),
  [\eprintmod[arXiv:astro-ph/0308461]].

\bibitem{Lasenby:2003ur}
A.~Lasenby and C.~Doran,
\newblock Phys. Rev. {\bf D71}, 063502 (2005),
  [\eprintmod[arXiv:astro-ph/0307311]].

\bibitem{Kesden:2003zm}
M.~H. Kesden, M.~Kamionkowski and A.~Cooray,
\newblock Phys. Rev. Lett. {\bf 91}, 221302 (2003),
  [\eprintmod[arXiv:astro-ph/0306597]].

\bibitem{Sinha:2005mn}
R.~Sinha and T.~Souradeep,
\newblock Phys. Rev. {\bf D74}, 043518 (2006),
  [\eprintmod[arXiv:astro-ph/0511808]].

\bibitem{Bridges:2005br}
M.~Bridges, A.~N. Lasenby and M.~P. Hobson,
\newblock Mon. Not. Roy. Astron. Soc. {\bf 369}, 1123 (2006),
  [\eprintmod[arXiv:astro-ph/0511573]].

\bibitem{Bridges:2006zm}
M.~Bridges, A.~N. Lasenby and M.~P. Hobson,
\newblock \mnras {\bf 381}, 68 (2007), [\eprintmod[arXiv:astro-ph/0607404]].

\bibitem{Boyanovsky:2006pm}
D.~Boyanovsky, H.~J. de~Vega and N.~G. Sanchez,
\newblock Phys. Rev. {\bf D74}, 123007 (2006),
  [\eprintmod[arXiv:astro-ph/0607487]].

\bibitem{Spergel:2006hy}
D.~N. Spergel {\em et~al.},
\newblock Astrophys. J. Suppl. {\bf 170}, 377 (2007),
  [\eprintmod[arXiv:astro-ph/0603449]].

\bibitem{Jain:2008dw}
R.~K. Jain, P.~Chingangbam, J.-O. Gong, L.~Sriramkumar and T.~Souradeep,
\newblock JCAP {\bf 0901}, 009 (2009), [\eprintmod[arXiv:0809.3915]].

\bibitem{Hu:1998kj}
W.~Hu,
\newblock Astrophys. J. {\bf 506}, 485 (1998),
  [\eprintmod[arXiv:astro-ph/9801234]].

\bibitem{Bean:2003fb}
R.~Bean and O.~Dore,
\newblock Phys. Rev. {\bf D69}, 083503 (2004),
  [\eprintmod[arXiv:astro-ph/0307100]].

\bibitem{dvali:00}
G.~R. Dvali, G.~Gabadadze and M.~Porrati,
\newblock Phys. Lett. {\bf B485}, 208 (2000),
  [\eprintmod[arXiv:hep-th/0005016]].

\bibitem{Fanetal08}
W.~Fang {\em et~al.},
\newblock Phys. Rev. {\bf D78}, 103509 (2008), [\eprintmod[arXiv:0808.2208]].

\bibitem{Dore:2003wp}
O.~Dore, G.~P. Holder and A.~Loeb,
\newblock Astrophys. J. {\bf 612}, 81 (2004),
  [\eprintmod[arXiv:astro-ph/0309281]].

\bibitem{Skordis:2004xr}
C.~Skordis and J.~Silk,
\newblock \eprintmod[arXiv:astro-ph/0402474].

\bibitem{GorHu04}
C.~Gordon and W.~Hu,
\newblock Phys. Rev. {\bf D70}, 083003 (2004),
  [\eprintmod[arXiv:astro-ph/0406496]].

\bibitem{Nolta:2008ih}
M.~R. Nolta {\em et~al.},
\newblock Astrophys. J. Suppl. {\bf 180}, 296 (2009),
  [\eprintmod[arXiv:0803.0593]].

\bibitem{Dunkley:2008ie}
J.~Dunkley {\em et~al.},
\newblock Astrophys. J. Suppl. {\bf 180}, 306 (2009),
  [\eprintmod[arXiv:0803.0586]].

\bibitem{Komatsu:2008hk}
E.~Komatsu {\em et~al.},
\newblock Astrophys. J. Suppl. {\bf 180}, 330 (2009),
  [\eprintmod[arXiv:0803.0547]].

\bibitem{Planck:2006uk}
{The Planck Collaboration},
\newblock \eprintmod[arXiv:astro-ph/0604069].

\bibitem{Lewis:1999bs}
A.~Lewis, A.~Challinor and A.~Lasenby,
\newblock Astrophys. J. {\bf 538}, 473 (2000),
  [\eprintmod[arXiv:astro-ph/9911177]].

\bibitem{Lewis:2002ah}
A.~Lewis and S.~Bridle,
\newblock Phys. Rev. {\bf D66}, 103511 (2002),
  [\eprintmod[arXiv:astro-ph/0205436]].

\bibitem{cosmomc_url}
\url{http://cosmologist.info/cosmomc/}.

\bibitem{wmaplike_url}
\url{http://lambda.gsfc.nasa.gov/}.

\bibitem{Mortonson:2008rx}
M.~J. {Mortonson} and W.~{Hu},
\newblock \apjl {\bf 686}, L53 (2008), [\eprintmod[arXiv:0804.2631]].

\bibitem{Hu:2003gh}
W.~Hu and G.~P. Holder,
\newblock Phys. Rev. {\bf D68}, 023001 (2003),
  [\eprintmod[arXiv:astro-ph/0303400]].

\bibitem{Mortonson:2007hq}
M.~J. Mortonson and W.~Hu,
\newblock Astrophys. J. {\bf 672}, 737 (2008), [\eprintmod[arXiv:0705.1132]].

\bibitem{cambrpc_url}
\url{http://background.uchicago.edu/camb_rpc/}.

\end{thebibliography}

\end{document}